\documentclass[5p,times,twocolumn]{elsarticle}
\biboptions{comma,sort&compress}
\usepackage[utf8]{inputenc}
\usepackage{bm}
\usepackage{hyperref} 
\usepackage[dvipsnames]{xcolor}
\usepackage{nicefrac}

\usepackage{amsmath,amsfonts,amsthm,amssymb}
\usepackage{nicefrac}
\usepackage{yfonts}

\hypersetup{
colorlinks = true,
linkcolor = blue,
anchorcolor = blue,
citecolor = blue,
filecolor = blue,
urlcolor = blue
}
\usepackage{ulem}
\graphicspath{{fig/}}

\def\spin{\,\textgoth{s:}}

\newcommand{\cA}{{\cal A}}
\newcommand{\pvv}{{\boldsymbol p}}
\newcommand{\sv}{{\boldsymbol s}}
\newcommand{\xv}{{\boldsymbol x}}

\journal{Physics Letters B}

\begin{document}

\begin{frontmatter}

\title{Polarization of spin-$\nicefrac{1}{2}$ particles with effective spacetime dependent masses}

\author[uj]{Samapan Bhadury}
\ead{samapan.bhadury@uj.edu.pl}

\author[ifj,bits]{Arpan Das}
\ead{arpan.das@ifj.edu.pl}
\ead{arpan.das@pilani.bits-pilani.ac.in}

\author[uj]{Wojciech Florkowski}
\ead{wojciech.florkowski@uj.edu.pl}

\author[iitg]{Gowthama K. K.}
\ead{k\_gowthama@iitgn.ac.in}

\author[ifj]{Radoslaw Ryblewski}
\ead{radoslaw.ryblewski@ifj.edu.pl}

\address[uj]{Institute of Theoretical Physics, Jagiellonian University, ul. St. \L ojasiewicza 11, 30-348 Krakow, Poland}
\address[ifj]{Institute of Nuclear Physics Polish Academy of Sciences, PL-31-342 Krakow, Poland}
\address[bits]{Department of Physics, Birla Institute of Technology and Science, Pilani, Rajasthan-333031, India}
\address[iitg]{Indian Institute of Technology Gandhinagar, Gandhinagar-382355, Gujarat, India}

\date{\today}

\begin{abstract} 
Semiclassical expansion of the Wigner function for spin-$\nicefrac{1}{2}$ fermions having an effective spacetime-dependent mass is used to analyze spin-polarization effects. The existing framework is reformulated to obtain a differential equation directly connecting the particle spin tensor with the effective mass. It reflects the conservation of the total angular momentum in a system. In general, we find that the gradients of mass act as a source of the spin polarization. Although this effect is absent for simple boost-invariant dynamics, an extension to non-boost-invariant systems displays a non-trivial dependence of the spin density on the mass indicating that the spin polarization effects may be intertwined with the phenomenon of chiral restoration. 
\end{abstract}


\end{frontmatter}


\section{Introduction}

The experimental results showing a non-zero spin polarization of the Lambda hyperons and vector mesons produced in relativistic heavy-ion collisions~\cite{STAR:2017ckg,STAR:2018gyt,STAR:2019erd,ALICE:2019onw,ALICE:2019aid,STAR:2020xbm,Kornas:2020qzi,STAR:2021beb,ALICE:2021pzu,STAR:2023ntw} triggered broad interest in the phenomena involving spin degrees of freedom~\cite{Becattini:2021lfq}. The measured global spin polarization along a direction perpendicular to the reaction plane is well explained within the models assuming thermalization of the spin degrees of freedom. On the other hand, other spin observables, in particular the azimuthal dependence of the longitudinal spin polarization, have yet to find a commonly accepted explanation. In this case, non-equilibrium phenomena are often invoked to explain the data~\cite{Becattini:2021iol,Fu:2021pok}.

Broad interest in spin physics and successes of relativistic hydrodynamics triggered the development of relativistic spin hydrodynamics \cite{Montenegro:2017rbu,Florkowski:2017ruc,Hidaka:2017auj,Florkowski:2018myy,Florkowski:2019qdp,Hattori:2019lfp,Fukushima:2020ucl,Shi:2020htn,Li:2020eon,Weickgenannt:2020aaf,Bhadury:2020puc,Peng:2021ago,Hu:2021pwh,Wang:2021ngp,Gallegos:2021bzp,Hongo:2021ona,She:2021lhe,Bhadury:2022ulr,Weickgenannt:2022zxs,Weickgenannt:2022jes,Weickgenannt:2022qvh,Cao:2022aku,Biswas:2023qsw}. Its dissipative extension forms a platform for addressing the problems of correct description of the longitudinal polarization~\cite{Becattini:2021iol,Fu:2021pok}. Moreover, spin hydrodynamics has been recently developed to include effects of the magnetic field~\cite{Bhadury:2022ulr}. 

Generally, we can conclude that important progress has been made in the analysis of the relativistic spin dynamics in hydrodynamic systems, and obviously also in the case where spin interacts (through magnetic moment) with the electromagnetic field. An example of the latter is the seminal Bargmann-Michel-Telegdi equation describing spin precession~\cite{PhysRevLett.2.435,Vasak:1987um}. On the other hand, the dynamics of spin in scalar fields attracted much less attention. In the present work, we intend to change this trend and indicate interesting phenomena that may take place in this case.

The coupling of spin to a scalar field may take place if such a field represents an effective particle mass. This type of interaction was studied in~\cite{Florkowski:1995ei} and more recently in~\cite{Wang:2021owk} \footnote{A fermion field can also acquire a spin-dependent contribution to its mass in the presence of a nonvanishing torsion of space–time as well as curvature-dependent effects~\cite{Struckmeier:2021rst}. However, in the present article, we do not discuss these scenarios.}. These two works are based on a semiclassical expansion of the Wigner function for spin-$\nicefrac{1}{2}$ fermions having an effective spacetime-dependent mass. An important difference between~\cite{Florkowski:1995ei} and~\cite{Wang:2021owk} is that the analysis in~\cite{Florkowski:1995ei} is restricted to the leading order of the semiclassical expansion, while~\cite{Wang:2021owk} addresses spin dynamics in higher orders, neglecting the leading order contribution. 

In this work, we reformulate the approach of Ref.~\cite{Florkowski:1995ei} to obtain a differential equation directly connecting the particle spin tensor with the effective mass. Interestingly, it reflects the conservation of the total angular momentum in a system. Our novel formulation of spin dynamics for particles with an effective spacetime-dependent mass allows for a straightforward and intuitive description of such systems. Our main finding is that the gradients of mass act as a source of the spin polarization, however, this effect is not manifested for simple boost-invariant dynamics.  
To examine some non-trivial features we studied a non-boost invariant system.

We use the convention $g_{\mu\nu} =  \hbox{diag}(+1,-1,-1,-1)$ and $\varepsilon^{0123} = -\varepsilon_{0123}=1$ for the metric tensor and totally anti-symmetric Levi-Civita symbol, respectively. Throughout the text, we use natural units.

\section{Spin-density dynamics}

Our starting point is the kinetic equation for the axial current phase-space density $\cA^\mu(x,k)$ originally derived in Ref.~\cite{Florkowski:1995ei}
\begin{eqnarray}
k^\alpha \partial_\alpha \cA^\mu
\!+\!M \partial_\alpha M \partial_{(k)}^\alpha  \cA^\mu
\!+\!\frac{\partial_\alpha M}{M} 
\left( 
k^\mu \cA^\alpha\!-\!k^\alpha \cA^\mu 
\right) = 0.
\label{eq:Hufner}
\end{eqnarray}
Here $M=M(x)$ is the in-medium space-time-dependent mass of particles, which is treated as externally given and plays the role of a background scalar field \cite{Gorenstein:1995vm, Romatschke:2011qp, Tinti:2016bav, Czajka:2017wdo}. Equation~(\ref{eq:Hufner}) has been derived within a semiclassical expansion for the Wigner function for particles with spin $\nicefrac{1}{2}$ and may be treated as an analog of the Bargmann-Michel-Telegdi equation that describes spin evolution in the electromagnetic field. The effects of particle collisions and thermalization are neglected here. 

In the leading order of the semiclassical expansion, one can use the following ansatz~\cite{Bhadury:2020puc}
\begin{eqnarray}
&& \cA^\mu(x,k) = 2 M \int dP dS \, s^\mu
\label{eq:A} \\ 
&& \times \left[
f^+(x,p,s) \delta^{(4)}(k-p)+
f^-(x,p,s) \delta^{(4)}(k+p)
\right], 
\nonumber 
\end{eqnarray}
where the functions $f^\pm(x,p,s)$ are distribution functions for particles ($f^+$) and antiparticles ($f^-$) in an extended phase space that includes space-time position $x^\mu =(t,\xv)$, momentum $p^\mu = (p^0, \pvv)$, and spin $s^\mu = (s^0, \sv)$. The four-momentum $p$ is always on the mass shell, $p^2~=~M^2(x)$, hence $p^0 = E_p = \sqrt{M^2(x) +\pvv^2}$. The form of Eq.~(\ref{eq:A}) implies that we can set $k = \pm p$ whenever $k$ directly multiplies the delta functions $\delta(k \mp p)$ in Eq.~(\ref{eq:A}). The latter property reflects the fact that Eq.~(\ref{eq:Hufner}) describes both particles and antiparticles.

The spin vector $s^\mu$ appearing in Eq.~(\ref{eq:A}) is defined by the expression~\cite{Bhadury:2020puc}
\begin{equation}
s^\mu = \frac{1}{2M} \, \varepsilon^{\mu\beta\gamma\delta} \, p_\beta \, s_{\gamma\delta},
\label{eq:s}
\end{equation}
where $s_{\gamma\delta}$ is the internal angular momentum tensor originally introduced by Mathisson~\cite{Mathisson:1937zz,2010GReGr..42.1011M}. We note that the momentum vector $p$ and the spin vector $s$ are orthogonal, $p \cdot s = 0$, and the inverse transformation to Eq.~(\ref{eq:s}) is $s^{\alpha\beta}= (1/M) \varepsilon^{\alpha\beta\gamma\delta} p_\gamma s_{\delta}$. The Lorentz invariant integration measures $dP$ and $dS$ are defined by the expressions $dP = d^3p/E_p$  and $dS= (M/\pi \spin)\, d^4s \, \delta(s \cdot s + \spin^2) \, \delta(p \cdot s)$, where for particles with spin $\nicefrac{1}{2}$ the variable $\spin$ denotes the value of the Casimir operator $\spin = (1/2) (1/2+1) = 3/4$. We note that Eq.~(\ref{eq:Hufner}) should be used along with the transversality condition~\cite{Florkowski:1995ei}
\begin{equation}
k_\mu \cA^\mu(x,k) = 0.  
\label{eq:Hufner_tr}
\end{equation}
Using Eq.~(\ref{eq:A}), one may check that this condition is consistent with Eq.~(\ref{eq:Hufner}). 

The axial current density is directly connected with the canonical spin tensor 
\begin{eqnarray}
S^{\lambda \mu \nu}_{\rm can}(x)\!=\!
\frac{1}{2} \varepsilon^{ \lambda \mu \nu \kappa} A_\kappa(x)
\!=\!
\frac{1}{2} \varepsilon^{\lambda \mu \nu \kappa} \int d^4k \, \cA_\kappa(x,k),
\label{eq:Scan}
\end{eqnarray}
which usually serves to analyze the spin polarization effects. In this work, we also use the GLW spin tensor defined by the formula~\cite{Bhadury:2020puc}
\begin{eqnarray}
&& S^{\lambda, \mu \nu}(x)\!=\!\int dP dS \, p^\lambda s^{\mu\nu} f(x,p,s), 
\label{eq:SGLW}
\end{eqnarray}
where $f(x,p,s) = f^+(x,p,s)\!+\!f^-(x,p,s)$. The acronym GLW refers to the names of de Groot, van Leeuwen, and van Weert who introduced this form of the spin tensor in their seminal textbook on the relativistic kinetic theory~\cite{DeGroot:1980dk} --- in contrast to the canonical spin tensor, the GLW version is conserved for free Dirac particles (with spacetime independent $M$).  Using Eq.~(\ref{eq:s}), we find the following relation
\begin{equation}
 S^{\lambda \mu \nu}_{\rm can} = S^{\lambda, \mu \nu} + S^{\mu, \nu \lambda} + S^{\nu, \lambda\mu}.
\label{eq:canGLW}
\end{equation}

Definitions introduced above may be used to transform Eq.~(\ref{eq:Hufner}) into an equation for the GLW spin tensor. This is achieved by the multiplication of Eq.~(\ref{eq:Hufner}) by  $k_\beta \varepsilon_\mu^{\,\,\,\beta \gamma \delta}$ and subsequent four-dimensional integration over $k$~\cite{Florkowski:2018ahw}. This leads us to the equation
\begin{equation}
\partial_\alpha S^{\alpha, \gamma \delta} = 
\frac{\partial_\alpha M}{M} \left(
S^{\gamma, \delta \alpha}
-S^{\delta, \gamma \alpha }
\right).
\label{eq:main}
\end{equation}


\section{Angular-momentum conservation law}

In order to clarify the physical interpretation of Eq.~(\ref{eq:main}), one can analyze the conservation of angular momentum in the discussed framework. Using Noether's Theorem one can obtain a general result that the divergence of the canonical spin tensor is equal to the difference of the asymmetric components of the canonical energy-momentum tensor
\begin{equation}
\partial_\lambda S^{\lambda \mu\nu}_{\rm can} =
T^{\nu\mu}_{(a) \,{\rm can}} - T^{\mu\nu}_{(a) \,{\rm can}}.
\label{eq:divS}
\end{equation}
Using the results for the semiclassical expansion of the Wigner function (for spin-$\nicefrac{1}{2}$ particles) one finds~\cite{Florkowski:2018ahw}
\begin{equation}
 T^{\mu\nu}_{(a) \,{\rm can}}(x) = \int d^4k \,k^\nu {\cal V}^\mu_{(1)}(x,k),  
\end{equation}
where ${\cal V}^\mu_{(1)}(x,k) = -(1/(2M)) \partial^\alpha {\cal S}_\alpha^{\,\,\mu}(x,k)$ is the first-order vector component of the Wigner function with
\begin{equation}
{\cal S}_{\alpha \mu}(x,k) = {1 \over M} \varepsilon_{\alpha\mu\rho\sigma} k^\rho {\cal A}^\sigma(x,k).
\label{eq:Salmu}
\end{equation}
With ${\cal A}^\sigma(x,k)$ defined by Eq.~(\ref{eq:A}), the angular-momentum conservation (\ref{eq:divS}) yields
\begin{equation}
M \partial_\alpha S^{\alpha \gamma\delta}_{\rm can} =  \partial_\alpha \left( M S^{\delta, \alpha\gamma} \right)
-  \partial_\alpha \left( M S^{\gamma, \alpha\delta} \right).
\label{eq:divSr}
\end{equation}
With the help of the relation~(\ref{eq:canGLW}), one can check  that Eq.~(\ref{eq:divSr}) is equivalent to Eq.~(\ref{eq:main}). Thus, our dynamic equation for the spin tensor $S^{\alpha, \gamma\delta}$ only reflects the conservation of the total angular momentum for particles with spacetime-dependent masses $M(x)$. 

We note that (\ref{eq:main}) implies that the GLW spin tensor is conserved in the case where $M$ is constant. This is expected as this version of the spin tensor has been constructed to exactly fulfill this requirement~\cite{DeGroot:1980dk}. In the present case, where the spacetime gradients of $M(x)$ are present, we observe that the right-hand side of Eq.~(\ref{eq:main}) acts as a source of the spin polarization. This requires, however, that there is already a non-zero spin polarization present in the system. 


\section{Transverse-polarization solution}
\label{sec:trans-sol}

Equation (\ref{eq:main}) can be easily solved if the dynamics of a spin-polarized system is essentially one-dimensional. This occurs in the early stages of heavy-ion collisions, where it is often assumed that the longitudinal expansion strongly dominates the transverse one. In this case, we assume 
\begin{equation}
f(x,p,s) = g(x,p,s) \delta(p_x) \delta(p_y)
\label{eq:deltas}
\end{equation}
and conclude, using Eq.~(\ref{eq:SGLW}), that $S^{1,\mu\nu}=S^{2,\mu\nu}=0$. Here we use the standard reference frame used in heavy-ion physics, with the $z$-axis coinciding with the beam axis, and the other two axes being perpendicular to it. Along with the assumption about the longitudinal expansion, we may assume that the effective mass as well as different components of the spin tensors are functions of the variables $t$ and $z$ only. 

\begin{figure}
\caption{\label{tp} Transverse-polarization case. Schematic view of a longitudinal expansion of the fireball formed in non-central heavy-ion collisions. The blue and red arrows represent the three-momentum vectors of particles and the direction of spin vectors, respectively. Boundary effects in the transverse plane are neglected.}
\includegraphics[width=0.5\textwidth]{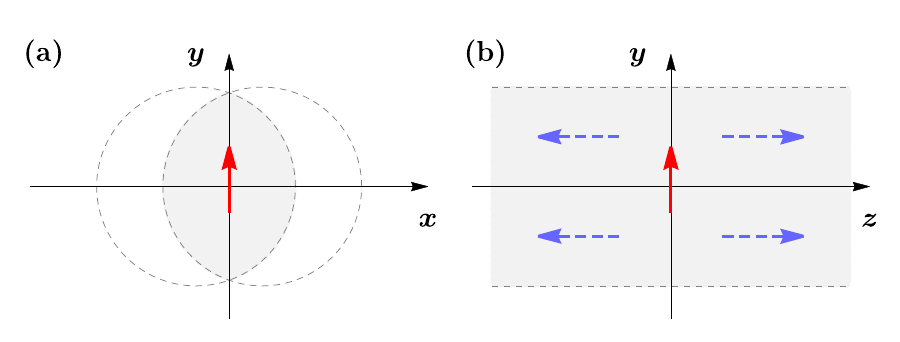}
\end{figure}

\begin{figure}
\caption{\label{lp} Longitudinal-polarization case. Notation the same as in Fig.~1. }
\includegraphics[width=0.5\textwidth]{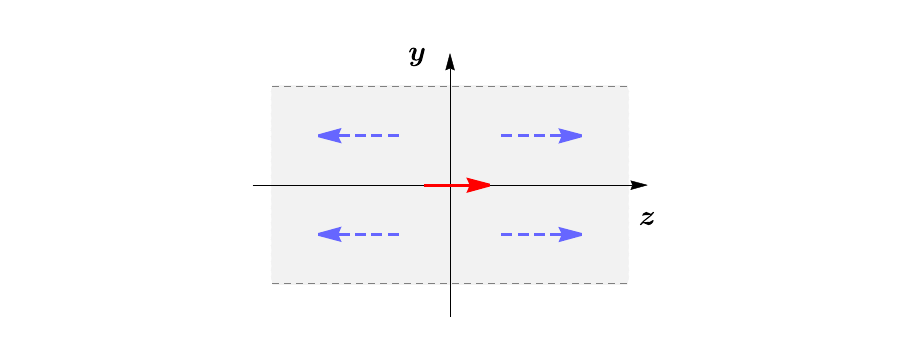}
\end{figure}

In non-central collisions, see Fig.~\ref{tp}, one commonly argues that some part of the initial orbital angular momentum is converted into a spin part (pointing along the $y$-axis) due to the spin-orbit coupling. To incorporate this idea into our description, we further consider the case where $g(x,p,s) = h(x,p,s) \delta(s_x) \delta(s_z)$.

Under the assumptions specified above, we deal with only four independent spin-tensor components: $S^{0,01} = -S^{0,10}$, $S^{3,01} = -S^{3,10}$, $S^{0,31} = -S^{0,13}$, $S^{3,31} = -S^{3,13}$, and the tensor structure of Eq.~(\ref{eq:main}) is reduced to two equations:
\begin{eqnarray}
\partial_0 S^{0,01} +  \partial_3 S^{3,01} 
&=& \frac{\partial_0 M}{M} S^{0,10}
+ \frac{\partial_3 M}{M} S^{0,13}
\nonumber \\
\partial_0 S^{0,31} +  \partial_3 S^{3,31} 
&=& \frac{\partial_0 M}{M} S^{3,10}
+ \frac{\partial_3 M}{M} S^{3,13}.
\label{eq:TP1}
\end{eqnarray}
These are two equations for four unknowns, which cannot be solved without imposing further constraints on the underlying distribution function. To make further progress we assume longitudinal boost invariance and use the form $S^{\lambda, \mu \nu} = \sigma(\tau) u^\lambda \varepsilon^{\mu \nu \alpha \beta} u_\alpha S_{y \beta}$ that is motivated by our form of $f(x,p,s)$. Here $\sigma$ is the spin density depending on the longitudinal proper time $\tau=\sqrt{t^2-z^2}$, $u^\mu = (t/\tau,0,0,z/\tau)$ is the Bjorken flow vector~\cite{Bjorken:1982qr}, and $S_y^\mu=(0,0,1,0)$ defines the transverse spin direction (along the $y$-axis). With $M$ depending on the proper time only, the two equations (\ref{eq:TP1})  reduce to a single differential equation 
\begin{equation}
\frac{d\sigma}{\, d\tau} +  \frac{\sigma}{\tau} = 0
\label{eq:TP2}
\end{equation}
with the solution
\begin{equation}
\sigma(\tau) = \sigma_0 \frac{\tau_0}{\tau},
\label{eq:sigma_sol_tau}
\end{equation}
where $\tau_0$ and $\sigma_0$ are integration constants.
It is easy to realize that Eq.~(\ref{eq:TP2}) is equivalent to the conservation law $\partial_\mu (\sigma u^\mu) = 0$ known from the Bjorken model~\cite{Bjorken:1982qr}. Hence, for the assumed spacetime expansion the change of the spin density is caused solely by the longitudinal expansion of the system --- it is decoupled from the changes of $M(\tau)$. 


\section{Longitudinal-polarization solution}
\label{sec:long-sol}

A complementary case to the one described above can be considered if the particles are longitudinally polarized, see Fig.~\ref{lp}. This case can be studied by assuming $g(x,p,s) = h(x,p,s) \delta(s_x) \delta(s_y)$. For a boost-invariant expansion with the only non-vanishing components $S^{0,12}(t,z)$ and $S^{3,12}(t,z)$, we can use the following parametrization
%
\begin{equation}
S^{\lambda,\mu\nu}(t,z) = u^\lambda \,\sigma(\tau) \epsilon^{\mu\nu\alpha\beta}u_{\alpha}S_{z\,\beta},
\label{eq:paramet}
\end{equation}
%
where $S_{z}^\mu = (0,0,0,1)$.
In this case, the spin evolution equation (\ref{eq:main}) has the same solution as that found in the transverse-polarization case, see Eqs.~(\ref{eq:TP2}) and (\ref{eq:sigma_sol_tau}).


\section{Solution for beyond boost-invariance}
\label{sec:nonboost-sol}

The results presented in Sec.~\ref{sec:trans-sol} and \ref{sec:long-sol} indicate that boost invariance strongly restricts the form of possible solutions of Eqs.~(\ref{eq:main}), leading to rather trivial solutions that express the spin density changes caused by longitudinal expansion. To find less trivial solutions, we abandon the assumption of boost invariance but still consider a one-dimensional system that expands along the $z$-axis. Since, in general, the system of Eqs.~(\ref{eq:main}) is underdetermined (includes 6 equations for 24 unknown components of the spin tensor) we further assume that the only non-zero components of the spin tensor are $S^{0,01}(t,z)$ and $S^{3,01}(t,z)$ (alternatively, we may select $S^{0,02}(t,z)$ and $S^{3,02}(t,z)$). Moreover, we assume that the mass $M$ depends only on the time coordinate, and $S^{3,01}(t,z) = v \, S^{0,01}(t,z) = v\, \sigma(t,z)$, where $v$ is a constant (satisfying the condition $|v|<1$).\footnote{In a way, our analysis here is similar to the analysis of the continuity equation of the form $\partial_0 \rho + \partial_3 j^{\,3} = 0$, with the assumption that the current $j^{\,3}$ is proportional to the density $\rho$, namely $j^{\,3} = v \rho$.}

In the considered case, Eqs.~\eqref{eq:main} are reduced to a single equation 
\begin{equation}
    \left(\frac{\partial}{\partial t} + v\,\frac{\partial}{\partial z}\right) \sigma (t,z) = - \frac{\partial \ln M(t)}{\partial t} \, \sigma (t,z). \label{eq:pde}
\end{equation}
Its formal solution has the form
\begin{equation}
\sigma(t,z) = \frac{M_0}{M(t)} \sigma_0\left(z - v (t-t_0)\right), 
\label{eq:nb-sol}
\end{equation}
where $M_0$ is the value of mass at some initial time $t=t_0$, while the function $\sigma_0$ defines the initial spatial profile of the spin density, $\sigma(t_0,z) =\sigma_0(z)$.

The solution (\ref{eq:nb-sol}) describes a spin polarized region characterized by the profile $\sigma_0(z)$ ``moving'' to the right with the velocity~$v$. The height of this profile is given by the ratio $M_0/M(t)$, hence the magnitudes of the spin polarization $\sigma$ and mass $M(t)$ are anticorrelated: for $M$ decreasing with time $\sigma$ grows and vice versa. This behavior indicates that the spin polarization phenomena may be connected with the effects of chiral symmetry restoration. Finally, we also note that using Eqs.~\eqref{eq:canGLW} and \eqref{eq:divSr}, some components of the canonical spin tensor (for example,  $S^{301}_{\rm can}$) are also found to have non-trivial dependence on the mass, $M(t)$. In this context see also our discussion in~\cite{Dey:2023hft}.


\section{Summary/Conclusions}

\begin{itemize}

\item In this work we have found that the gradients of an effective mass can act as a source of the spin polarization (quantified by the increase of the divergence of the GLW spin tensor). Although this effect has been found to be absent for Bjorken expansion (simple one-dimensional boost invariant flow without transverse expansion), a more generic one-dimensional expansion indicates the correlation (anti-correlation) between the spin polarization and the mass of the constituents of the medium.

\item  We stress that the right-hand side of Eq.~(\ref{eq:main}) is zero for a vanishing spin tensor. Hence, it acts as a source/sink of the spin polarization only if there is already a non-negligible spin polarization in the system. This is similar in structure to the hydrodynamic equations which describe (for example) changes in energy density and pressure and cannot explain the initial values of these quantities.

\item We have treated the effective masses of particles as given externally. In future works, one may study coupled equations where $M(x)$ is determined by the spacetime (scalar) density of particles. In particular, one may consider the Nambu--Jona-Lasinio (NJL) type of interaction and use the NJL gap equation to self-consistently determine the function $M(x)$.

\item  We note that the total spin included in the space-time volume $\Sigma$ characterized by the volume element $d\Sigma_\mu$ is
\begin{equation}
\Delta S^{\alpha \beta} = 
\int_\Sigma  d\Sigma_\mu \, S^{\mu, \alpha\beta}.
\end{equation}
For the boost-invariant cases discussed above $d\Sigma_\mu = u_\mu \, dx dy d\eta$, with $\eta$ being the spacetime rapidity. For possible comparisons with the data, it is also recommend to switch finally to the canonical spin tensor as it has been argued recently in~\cite{Dey:2023hft}.

\end{itemize}


\section{Acknowledgements}

We thank Sourav Dey for illuminating discussions. This work was supported in part by the Polish National Science Centre Grants Nos. 2018/30/E/ST2/00432, and 2022/47/B/ST2/01372. GKK acknowledges the hospitality of the Institute of Nuclear Physics PAN, Krak\'ow, Poland, and the Indian Institute of Technology, Gandhinagar (IIT GN) for the Overseas Research Experience Fellowship to visit INP PAN. SB kindly acknowledges the support of the Faculty of Physics, Astronomy and Applied Computer Science, Jagiellonian University Grant No. LM/17/BS. AD would like to acknowledge the New Faculty Seed Grant (NFSG), NFSG/PIL/2024/P3825, provided by the Birla Institute of Technology and Science, Pilani, India.
%

\bibliographystyle{elsarticle-num}
\bibliography{ref}

\end{document}